\documentclass[manyauthors,nocleardouble]{cernphprep}
\usepackage{amsmath,amssymb,amsfonts,amsbsy}
\usepackage{bm} 
\usepackage{cite}
\usepackage{wrapfig}                                                         %
\newcommand{\id}{\mathrm{d}}
\usepackage{bm} 
\usepackage{color}                                                       

\begin{document}

\begin{titlepage}
\PHnumber{2016--245.R5}
\PHdate{\today}

\title{Azimuthal asymmetries of charged hadrons produced in
high-energy muon scattering off longitudinally polarised
deuterons}

\Collaboration{The COMPASS Collaboration}
\ShortAuthor{The COMPASS Collaboration}
\ShortTitle{Azimuthal asymmetries \ldots}

\begin{abstract}
Single hadron azimuthal asymmetries of positive and negative hadrons produced in muon semi-inclusive deep inelastic scattering off longitudinally polarised deuterons are determined using the 2006 COMPASS data and also combined all deuteron COMPASS data. For each hadron charge, the dependence of the azimuthal asymmetry on the hadron azimuthal angle $\phi$ is obtained by means of a five-parameter fitting function that besides a $\phi$-independent term includes four modulations
predicted by theory: $\sin\phi$, $\sin 2 \phi$, $\sin 3\phi$ and $\cos\phi$. The amplitudes of the five terms have been extracted, first, for the hadrons in the whole available kinematic region.
In further fits, performed for hadrons from a restricted
kinematic region, the $\phi$-dependence is determined as a
function of one of three variables (Bjorken-$x$,
fractional energy of virtual photon taken by the outgoing hadron and hadron transverse momentum), while  disregarding the others. Except the $\phi$-independent term, all the modulation amplitudes are very small, and no clear kinematic dependence could be observed within experimental uncertainties.

\end{abstract}

\vspace*{60pt}
\begin{flushleft}
\end{flushleft}

\vfill
\Submitted{(To be submitted to the European Physical Journal C)}
\end{titlepage}

{\pagestyle{empty}
%
%
\section*{The COMPASS Collaboration}
\label{app:collab}
\renewcommand\labelenumi{\textsuperscript{\theenumi}~}
\renewcommand\theenumi{\arabic{enumi}}
\begin{flushleft}
C.~Adolph\Irefn{erlangen},
M.~Aghasyan\Irefn{triest_i},
R.~Akhunzyanov\Irefn{dubna}, 
M.G.~Alexeev\Irefn{turin_u},
G.D.~Alexeev\Irefn{dubna}, 
A.~Amoroso\Irefnn{turin_u}{turin_i},
V.~Andrieux\Irefnn{illinois}{saclay},
N.V.~Anfimov\Irefn{dubna}, 
V.~Anosov\Irefn{dubna}, 
K.~Augsten\Irefnn{dubna}{praguectu}
% ...
\end{flushleft}
%
%
\begin{Authlist}
\item \Idef{turin_p}{University of Eastern Piedmont, 15100 Alessandria, Italy}
\item \Idef{aveiro}{University of Aveiro, Department of Physics, 3810-193 Aveiro, Portugal}
\item \Idef{bochum}{Universit\"at Bochum, Institut f\"ur Experimentalphysik, 44780 Bochum, Germany\Arefs{l}\Arefs{s}}
\item \Idef{bonniskp}{Universit\"at Bonn, Helmholtz-Institut f\"ur  Strahlen- und Kernphysik, 53115 Bonn, Germany\Arefs{l}}
\item \Idef{bonnpi}{Universit\"at Bonn, Physikalisches Institut, 53115 Bonn, Germany\Arefs{l}}
\item \Idef{brno}{Institute of Scientific Instruments, AS CR, 61264 Brno, Czech Republic\Arefs{m}}
\item \Idef{calcutta}{Matrivani Institute of Experimental Research \& Education, Calcutta-700 030, India\Arefs{n}}
\item \Idef{dubna}{Joint Institute for Nuclear Research, 141980 Dubna, Moscow region, Russia\Arefs{o}}
\item \Idef{erlangen}{Universit\"at Erlangen--N\"urnberg, Physikalisches Institut, 91054 Erlangen, Germany\Arefs{l}}
\item \Idef{freiburg}{Universit\"at Freiburg, Physikalisches Institut, 79104 Freiburg, Germany\Arefs{l}\Arefs{s}}
\item \Idef{cern}{CERN, 1211 Geneva 23, Switzerland}
\item \Idef{liberec}{Technical University in Liberec, 46117 Liberec, Czech Republic\Arefs{m}}
\item \Idef{lisbon}{LIP, 1000-149 Lisbon, Portugal\Arefs{p}}
\item \Idef{mainz}{Universit\"at Mainz, Institut f\"ur Kernphysik, 55099 Mainz, Germany\Arefs{l}}
\item \Idef{miyazaki}{University of Miyazaki, Miyazaki 889-2192, Japan\Arefs{q}}
\item \Idef{moscowlpi}{Lebedev Physical Institute, 119991 Moscow, Russia}
\item \Idef{munichtu}{Technische Universit\"at M\"unchen, Physik Department, 85748 Garching, Germany\Arefs{l}\Arefs{r}}
\item \Idef{nagoya}{Nagoya University, 464 Nagoya, Japan\Arefs{q}}
\item \Idef{praguecu}{Charles University in Prague, Faculty of Mathematics and Physics, 18000 Prague, Czech Republic\Arefs{m}}
\item \Idef{praguectu}{Czech Technical University in Prague, 16636 Prague, Czech Republic\Arefs{m}}
\item \Idef{protvino}{State Scientific Center Institute for High Energy Physics of National Research Center `Kurchatov Institute', 142281 Protvino, Russia}
\item \Idef{saclay}{IRFU, CEA, Universit\'e Paris-Saclay, 91191 Gif-sur-Yvette, France\Arefs{s}}
\item \Idef{taipei}{Academia Sinica, Institute of Physics, Taipei 11529, Taiwan}
\item \Idef{telaviv}{Tel Aviv University, School of Physics and Astronomy, 69978 Tel Aviv, Israel\Arefs{t}}
\item \Idef{triest_u}{University of Trieste, Department of Physics, 34127 Trieste, Italy}
\item \Idef{triest_i}{Trieste Section of INFN, 34127 Trieste, Italy}
\item \Idef{triest_ictp}{Abdus Salam ICTP, 34151 Trieste, Italy}
\item \Idef{turin_u}{University of Turin, Department of Physics, 10125 Turin, Italy}
\item \Idef{turin_i}{Torino Section of INFN, 10125 Turin, Italy}
\item \Idef{illinois}{University of Illinois at Urbana-Champaign, Department of Physics, Urbana, IL 61801-3080, USA}
\item \Idef{warsaw}{National Centre for Nuclear Research, 00-681 Warsaw, Poland\Arefs{u} }
\item \Idef{warsawu}{University of Warsaw, Faculty of Physics, 02-093 Warsaw, Poland\Arefs{u} }
\item \Idef{warsawtu}{Warsaw University of Technology, Institute of Radioelectronics, 00-665 Warsaw, Poland\Arefs{u} }
\item \Idef{yamagata}{Yamagata University, Yamagata 992-8510, Japan\Arefs{q} }
\end{Authlist}
%
%
\renewcommand\theenumi{\alph{enumi}}
\begin{Authlist}
\item [{\makebox[2mm][l]{\textsuperscript{*}}}] Deceased
\item [{\makebox[2mm][l]{\textsuperscript{\#}}}] Corresponding authors
\item \Adef{a}{Also at Instituto Superior T\'ecnico, Universidade de Lisboa, Lisbon, Portugal}
\item \Adef{b}{Also at Department of Physics, Pusan National University, Busan 609-735, Republic of Korea and at Physics Department, Brookhaven National Laboratory, Upton, NY 11973, USA}
\item \Adef{r}{Supported by the DFG cluster of excellence `Origin and Structure of the Universe' (www.universe-cluster.de)}
\item \Adef{d}{Also at Chubu University, Kasugai, Aichi 487-8501, Japan\Arefs{q}}
\item \Adef{x}{Also at Department of Physics, National Central University, 300 Jhongda Road, Jhongli 32001, Taiwan}
\item \Adef{e}{Also at KEK, 1-1 Oho, Tsukuba, Ibaraki 305-0801, Japan}
\item \Adef{g}{Also at Moscow Institute of Physics and Technology, Moscow Region, 141700, Russia}
\item \Adef{v}{Supported by Presidential grant NSh--999.2014.2}
\item \Adef{h}{Present address: RWTH Aachen University, III.\ Physikalisches Institut, 52056 Aachen, Germany}
\item \Adef{y}{Also at Department of Physics, National Kaohsiung Normal University, Kaohsiung County 824, Taiwan}
\item \Adef{i}{Present address: Uppsala University, Box 516, 75120 Uppsala, Sweden}
\item \Adef{c}{Supported by the DFG Research Training Group Programmes 1102 and 2044} 
%
%
\item \Adef{l}{Supported by the German Bundesministerium f\"ur Bildung und Forschung}
\item \Adef{s}{Supported by EU FP7 (HadronPhysics3, Grant Agreement number 283286)}
\item \Adef{m}{Supported by Czech Republic MEYS Grant LG13031}
\item \Adef{n}{Supported by SAIL (CSR), Govt.\ of India}
\item \Adef{o}{Supported by CERN-RFBR Grant 12-02-91500}
\item \Adef{p}{\raggedright Supported by the Portuguese FCT - Funda\c{c}\~{a}o para a Ci\^{e}ncia e Tecnologia, COMPETE and QREN,
 Grants CERN/FP 109323/2009, 116376/2010, 123600/2011 and CERN/FIS-NUC/0017/2015}
\item \Adef{q}{Supported by the MEXT and the JSPS under the Grants No.18002006, No.20540299 and No.18540281; Daiko Foundation and Yamada Foundation}
\item \Adef{t}{Supported by the Israel Academy of Sciences and Humanities}
\item \Adef{u}{Supported by the Polish NCN Grant 2015/18/M/ST2/00550}
\end{Authlist}

\clearpage
}

\setcounter{page}{1}
\section{Introduction}

Measurements of Semi-Inclusive Deep-Inelastic Scattering (SIDIS)
\begin{equation}
\label{eq1} {\mu}+{N}\to \mu'+ nh + X, \qquad n=1,2, ...
\end{equation}
of high-energy polarised muons $\mu$ off nucleons $N$
in the initial state  and scattered muons $\mu'$, $n$ measured
hadrons $h$ and unobserved particles $X$ in the final state are
sensitive to the spin-dependent Parton Distribution Functions
(PDFs) of nucleons. The SIDIS cross section depends, in
particular, on the azimuthal angle of each produced and measured hadron (see
e.g.\ Ref.~\cite{Kotzinian:1995cz}), which leads to azimuthal
asymmetries related to convolutions of the nucleon
Transverse-Momentum-Dependent (TMD) PDFs and parton-to-hadron Fragmentation
Functions (FFs). These asymmetries can appear in SIDIS off unpolarised,
longitudinally or transversely polarised nucleons.

The TMD PDFs were studied in a number of experiments. The short overview of earlier results obtained by the HERMES, CLAS and COMPASS collaborations on azimuthal asymmetries in SIDIS production of charged hadrons was given in Ref. \cite{Alekseev:2010dm}. The COMPASS collaboration has published results on asymmetries off unpolarised $^6$LiD (referred to as "deuteron") target \cite{Adolph:2014pwc}, transversely polarised deuterons \cite{Ageev:2006da} and transversely polarised NH$_3$ (referred to as  "proton") target \cite{Adolph:2012sp}. The common analysis of transversely polarised deuteron and proton data is included in Ref. \cite{Adolph:2012sp} also. The updated overview of the TMD PDFs including the COMPASS results can be found in Ref. \cite{Avakian:2016rst}. The COMPASS results on azimuthal asymmetries off longitudinally polarised deuterons based on the data collected in 2002, 2003 and 2004 were published in Ref. \cite{Alekseev:2010dm} for so called
``integrated" asymmetries and asymmetries as functions
of kinematic variables extracted in the restricted kinematic region.
Similar data have been collected in 2006 but not published yet. The results on the integrated asymmetries for 2006 and for the combined 2002 -- 2006 data which are presented
in this Paper are extracted using hadrons
from the whole available kinematic region, at variance with Ref. \cite{Alekseev:2010dm}, while asymmetries as functions of the kinematic variables are extracted using the hadrons from
a restricted kinematic region, similar to Ref. \cite{Alekseev:2010dm}.

The Paper is organised as follows. The SIDIS kinematics, basic
formulae and a brief theoretical overview are given in Section 2.
The analysis of the 2006 data is described in Section 3. The
results on the asymmetries of the combined 2002 -- 2006 data are
presented in Section 4. Systematic uncertainties are discussed in
Section 5 and conclusions are given in Section 6.

\section{Theoretical framework}
\begin{figure}[h!]
\centering 
{\rotatebox{-90}{
\includegraphics[width=.3\textwidth]{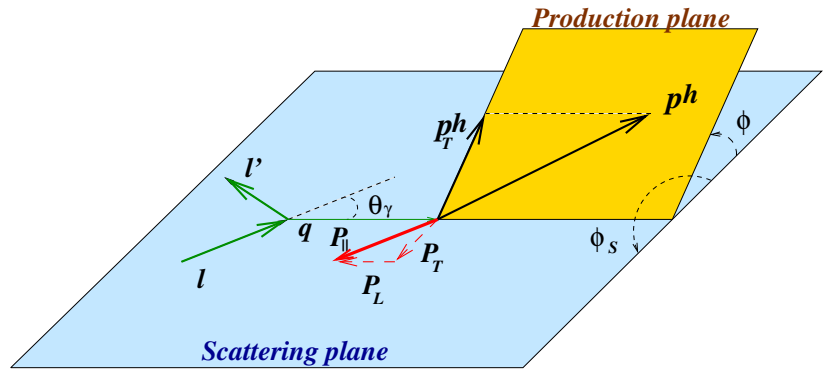}}}
\vskip+2mm
\caption{\label{fig1}\footnotesize%
 The SIDIS kinematics shown for target deuteron
polarisation ${\bm P}_\|$ antiparallel to the beam direction.}
\end{figure}

The SIDIS kinematics is illustrated in Fig.~\ref{fig1}.

The 4-momenta of the incident and scattered muons are denoted by $l$
and $l'$, respectively. The 4-momentum of the virtual photon is
given by $q=l-l'$ with $Q^2=-q^2$. The angle of the momentum
vector $\bm{q}$ of the virtual photon with respect to the
incident muon is denoted by $\theta_\gamma$. The vectors $\bm p^h$ and $\bm P_\|$ denote the
hadron momentum and the longitudinal target deuteron polarisation,
respectively. Their transverse components ${\bm p}_T^h$ and ${\bm
P_T}$ are defined with respect to the virtual-photon momentum.
The longitudinal component $|{\bm P_L}|= |{\bm
P_\|}|\cos\theta_\gamma$ is approximately equal to $|\bm{P_\|}|$
due to the smallness of the angle $\theta_\gamma$. The small
transverse component is equal to $\left|{\bm P_T}\right|=|{\bm
P_{\|}}|\sin\theta_\gamma$ where $\sin\theta_\gamma\approx
2({Mx}/{Q})\sqrt{1 - y}$. Here, $M$ is the nucleon mass and
$y= (qp)/(pl)$ is the fractional energy of the virtual photon,
where $p$ is the 4-momentum of the target nucleon. The angle
$\phi$ denotes the azimuthal angle between the lepton scattering
plane and the hadron production plane, while $\phi_S$ denotes the angle
of the deuteron polarisation vector with respect to the  scattering plane:
$\phi_S=0^\circ$  or $180^\circ$  for deuteron polarisation parallel or
antiparallel to the beam direction, respectively. Furthermore, the Bjorken
variable, $x_{Bj}\equiv x =Q^2/(2pq)$, the fraction of the
virtual-photon energy taken by a hadron, $z=(pp^h)/(pq)$,
the transverse momentum of a hadron, $p^h_T$, and the
invariant mass of the photon-nucleon system, $W^2=(p+q)^2$,
that, together with $Q^2 > 1$ (GeV/$c$)$^2$ and $0 <y< 1$, characterise SIDIS under study.

The general expression for the differential SIDIS cross section
(see Ref.~\cite{Kotzinian:1995cz} and references therein) is a
linear function of the incident muon polarisation $P_\mu$ and of
the longitudinal and transverse components \boldmath${\bm P}_L$  and
\boldmath${\bm P}_T$ of the target deuteron polarisation $\bm{P}_\|$:%
\vskip-3mm
\begin{equation}
\label{eq0} \id\sigma=\id\sigma_{00}+{\bm P_\mu} \id\sigma_{L0}+{\bm
P_L}\left( {\id\sigma_{0L} +{\bm P_\mu} \id\sigma_{LL}}\right)+{\bm
P_T} \left({\id\sigma_{0T}+{\bm P_\mu} \id\sigma_{LT}}\right) .
\end{equation}
\vskip-1mm%
Here, the first (second) subscript of the partial cross sections
refers to the beam (target) polarisation: $0,\ L$ or $T$ denote
unpolarised, longitudinally or transversely polarised.

The azimuthal asymmetries of charged hadron production
$a_{h^\pm}(\phi)$
are defined as follows:
\vskip-2mm
\begin{equation}
a_{h^\pm}(\phi)= \frac{\id\sigma^{\leftarrow\Rightarrow}-
\id\sigma^{\leftarrow\Leftarrow}}
{|P_L|(\id\sigma^{\leftarrow\Rightarrow}+\id\sigma^{\leftarrow\Leftarrow})}~,
\label{eq2}
\end{equation}
\vskip-1mm
where all cross sections are  functions of the angle $\phi$. The
Eq.~(\ref{eq2}) represents a definition of the experimentally measured asymmetries common for this Paper and for Refs. \cite{Alekseev:2010dm,Ageev:2005gh}.  The first (second) superscript
denotes the beam (target) spin orientation. The
symbol $\leftarrow$ denotes the incident muon spin orientation
that, in the case of a positive charge of the incident muons, is
mainly opposite to the beam direction. For the CERN muon beam, the average value of $|{\bm P_\mu}|$ is equal to $0.8$. The beam polarisation does not enter in the definition of measured asymmetries. The symbols $\Rightarrow$ and $\Leftarrow$ denote the target deuteron spin orientations (polarisations) the first of which is parallel, considered further as positive $(+)$, and the second one is antiparallel $(-)$ to the beam direction (see Section 3.1).

Substituting the general expression for $\id\sigma$ (Eq. (\ref{eq0})) in the cross sections of the  Eq. (\ref{eq2}), one can obtain the expected contributions of the partial cross sections to the azimuthal asymmetries. As the result, when taking into account the signs of the target polarizations, one can see that only four partial cross sections contribute to the numerator of Eq. (\ref{eq2}) and two to its denominator. In the numerator, we expect to have contributions from $\id\sigma_{0L}$, $P_\mu\id\sigma_{LL}$ and $\tan\theta_\gamma(\id\sigma_{0T}+P_\mu\id\sigma_{LT})$, while in the denominator from $\id\sigma_{00}$ and $P_\mu \id\sigma_{L0}$. The explicit expression for these partial cross sections in terms of the PDFs and their dependences on the hadron azimuthal angle have been given in Ref. \cite{Alekseev:2010dm} and briefly commented below.

\begin{figure}[h!]
\centering
\includegraphics[width=.4\textwidth]{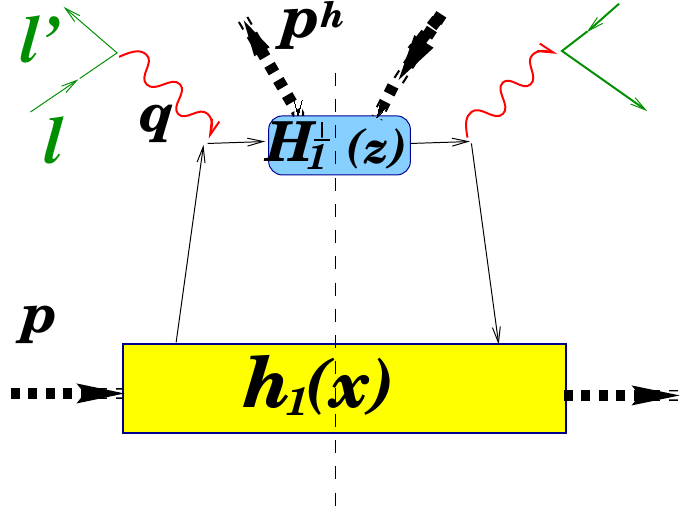}
\vskip-1mm
\caption{\label{fig1b}\footnotesize%
 The diagram  describing the contribution to the SIDIS
cross section of PDF $h_1(x)$ convoluted with 
FF $H_1^\bot(z)$.}
\end{figure}

Following phenomenological considerations based on the QCD parton
model of the nucleon and SIDIS in one-photon exchange
approximation,
the squared modulus of the matrix element, defining the cross sections,
is represented by a number of diagrams. As an example\footnote{In
this Paper we follow the Amsterdam notations for PDFs and PFFs,
see e.g.\ Ref.~\cite{Kotzinian:1995cz}.}, the diagram accounting
for the contribution to the SIDIS cross section of the chiral-odd
transversity PDF $h_{1}(x)$ convoluted with the chiral-odd
Collins FF $H_1^\bot(z)$ is shown in Fig.~\ref{fig1b}.
The diagram contributes to the asymmetry Eq.~(\ref{eq2}) via the term
$\id\sigma_{0T}$. Other PDFs convoluted with corresponding FFs can
also contribute to the cross section of spinless or unpolarised hadron production
off longitudinally polarised deuterons and their expected azimuthal modulations.
As motivated in Ref. \cite{Alekseev:2010dm}, out of predicted terms the $\phi$-independent term $a^0_{h^\pm}$ and four
modulation terms up to the order of $M/Q$ are retained for the analysis:
\vskip-2mm
\begin{equation}
\label{eq3}
a_{h^\pm}(\phi)=a_{h^\pm}^{0} + a_{h^\pm
}^{\sin\phi}\sin\phi+ a_{h^\pm}^{\sin2\phi}\sin2\phi +
a_{h^\pm}^{\sin3\phi}\sin3\phi + a_{h^\pm}^{\cos\phi}\cos\phi .
\end{equation}
\vskip-1mm
The sign and the amplitude of each modulation is a subject of the
$a_{h^\pm}(\phi)$ data analysis (see Section 3.6).
In Eq. (\ref{eq3}), the term
$a_{h^\pm}^{0}$ is related to the well known helicity PDF $g_{1L}$ contributing to asymmetries via $d\sigma_{LL}$. The two terms
with amplitudes $a_{h^\pm}^{\sin2\phi}$ and $a_{h^\pm}^{\sin\phi}$
reported in Ref. \cite{Airapetian:2002mf} are
related to the worm-gear-L PDF $h_{1L}^\bot$,
and to the PDFs $h_L$ and $f_L$, respectively, contributing to the
asymmetries via $d\sigma_{0L}$.
The  transversity PDF $h_1$ and Sivers PDF $f_{1T}^\bot$ can also contribute to the term
with amplitude $a_{h^\pm }^{\sin\phi}$ via $d\sigma_{0T}$ with a small factor $\tan\theta_\gamma$. Other
terms in Eq. (\ref{eq3}), not seen yet in experiments with longitudinally polarised deuterons,
are the terms with amplitudes $a_{h^\pm}^{\sin3\phi}$ and $a_{h^\pm}^{\cos\phi}$. They are related to the pretzelosity PDF $h_{1T}^\bot$  and the worm-gear-T PDF $g_{1T}$
contributing to asymmetries via $d\sigma_{0T}$ and $d\sigma_{LT}$, respectively, suppressed by the small factor $\tan\theta_\gamma$. The COMPASS
results \cite{Alekseev:2010dm} obtained from  the 2002 -- 2004
data suggested some indications for a possible $x$-dependence of terms with amplitudes $a_{h^\pm}^{\sin2\phi}$ and
$a_{h^\pm}^{\cos\phi}$. The
contribution of the term with amplitude $a_{h^\pm}^{\cos2\phi}$ which could have appeared from $\id\sigma_{00}$ in the denominator of Eq.~(\ref{eq2}) is disregarded. This amplitude is expected
\cite{arneodo,Airapetian:2012yg} to be of the
order of 0.1 and would enter Eq.~(\ref{eq3}) with the factor
$a^0_{h^\pm}$, that is of the order of $10^{-3}$ for integrated asymmetries (see
Table 2), or with $a_{h^\pm }^0 (x)\leq 0.05$ for asymmetries as functions of kinematic variables  (see Fig. \ref{fig4}).
This is beyond our experimental accuracy. The same comments apply to possible contributions
of terms with amplitudes $a_{h^\pm}^{\cos\phi}$ and $a_{h^\pm}^{\sin\phi}$ which also could originate from $\id\sigma_{00}$ and $\id\sigma_{L0}$ of the denominator of Eq.~(\ref{eq2}),
respectively. The negligible impact of the disregarded modulations on the amplitudes in Eq.~(\ref{eq3}) is confirmed by the 2006 data (see Section 3.6). All modulation amplitudes obtained in this Paper
refer to the average value of the beam polarisation equal to $-80\%$.

The aim of this study is to continue searches for possible
modulations in $a_{h^\pm}(\phi)$ as manifestation of TMD PDFs
describing the nucleons in the deuteron and to investigate the
$x$, $z$ and $p_T^h$ dependences of the corresponding modulation
amplitudes. For these purposes, we used first the 2006 deuteron data and then the combination
of all 2002 -- 2006 COMPASS deuteron data
with longitudinal target polarisation.

\section{The 2006 data analysis}

\subsection{Experimental set-up}
The COMPASS set-up is a two-stage forward spectrometer with the
world's largest polarised target and various types of tracking and particle
identification detectors (PID) in front and behind of two large-aperture magnets SM1 and SM2. These detectors provide data for reconstruction of corresponding tracks. The spectrometer was operated in the high energy
(160~GeV) muon beam at CERN. Its initial configuration
(see Ref.~\cite{Abbon:2007pq}) was used for
data taking in 2002 -- 2004. During the long accelerator shutdown in
2005,  the set-up was modified (see Ref.~\cite{Alekseev:2009ac}).
The major modifications influencing the present analysis were as
follows: (i) the replacement of the two 60~cm long target cells
(denoted as $U$ and $D$) by three cells $U,\ M$ and $D$ of
lengths 30~cm, 60~cm and 30~cm, (ii) the replacement of
the target solenoid magnet by the new one with a wider aperture
and (iii) the installation of the electromagnetic calorimeter
ECAL1 in front of the hadron calorimeter HCAL1.
The ECAL1 is not used in the analysis because it was not fully operational yet in 2006 and partially acted as a hadron absorber. These modifications of the apparatus where aimed at further reduction of systematic uncertainties, enlargement of the spectrometer acceptance  and  improvement of $e/\gamma$ ~PID capabilities. These modifications have required reconsidering the Ref.
\cite{Alekseev:2010dm} methods of data stability tests and asymmetry
calculations (see Sections 3.3 and 3.6.)

The data in 2006 were taken in two groups of periods. Each group
is characterised by its initial set of polarisations in the
target cells which are obtained by using different frequencies
of the microwave field to polarise the target material (deuterons) in
different cells at the certain direction of the target magnet solenoid field. The solenoid fields holds the polarisation. The field direction is denoted as $f=+$, if it coincides with the beam direction, or $f=-$, if opposite. The first group of the periods is denoted by G1 and other one by
G2. Each period includes a certain number of intervals of
continuous data taking (referred to as runs). The {G1} data
taking periods started with the initial setting of positive
deuteron polarisation in target cells {\it U} and {\it D} and the
negative one in cell {\it M}, both corresponding to $f = +$. After
taking some number of runs, the field was reversed to $f=-$ causing the reversal of the target cell polarisations, so that the data were taken with
opposite deuteron polarisations in the cells. The periodic
reversal of polarisations continued up to the end of G1 periods.
Within the periods, the cell polarisations,  needed for
asymmetry calculations (see Section 3.5), were measured for each
run in order to make sure that they are stable at the level of
about 55\%. If polarisations dropped below this limit, they were
restored by the microwave field before the beginning of the next
period. For {G2} periods, the procedure was analogous but the
initial setting of polarisation in the cells was opposite to the
one in {G1} at the same field $f=+$.
%
The periodic reversal of the cell polarisations within each group
of periods was used to estimate a possible time-dependent
systematics of the data. The change of the initial setting of the
cell polarisations was used to estimate a possible systematic
change of the spectrometer acceptance due to superposition of the
solenoid field and the field of SM1. If there is no such
systematic change, the acceptance in G1 and G2 periods must be the
same  for stable performance of the spectrometer.
\subsection{Selection of SIDIS events and hadrons}
Let us call as "SIDIS event" an event determined by Eq.~(\ref{eq1})
and reconstructed with tracks using the data recorded by
the tracking and PID detectors.

The overall statistics of 2006 is about $44.6\times 10^6$ of
preselected candidates for inclusive DIS and SIDIS event
with $Q^2>1$~(GeV/$c)^2$. The sample was obtained after rejection of runs that did not pass the data stability tests (see Section 3.3) and events that did not pass the reconstruction tests. The latter ones were rejected
if the $Z$-coordinate (along the beam) of the interaction point (vertex) was
determined with an uncertainty larger than $3\sigma$ of average
which varied within 1.5-2 cm for different target cells.

The selection of SIDIS events from the preselected sample was
done as described in Ref.~\cite{Alekseev:2010dm}. For each SIDIS event, a
reconstructed vertex with incident ($\mu$) and scattered ($\mu^\prime$) muons
and one or more additional tracks were required. Trajectories of the incident muons were required to traverse all target cells in order to have the same beam intensity for each of
them.
The track crossing more than 30 radiative lengths along the reconstructed trajectory was associated with $\mu^\prime$.
The cuts were applied on the quality of the reconstructed tracks forming vertices, the effective lengths of the target cells
(28~cm, 56~cm, 28~cm), the momentum of incident muons (140~GeV/$c$ $-$ 180~GeV/$c$), the fractional energy carried by
all tracks from the event $(z < 1)$ and the fractional
virtual-photon energy $(0.1< y < 0.9)$. About $36.6\times10^6$ SIDIS event candidates remained after cuts.

The distribution of  track multiplicities per SIDIS candidates peaks at four. These tracks include scattered $\mu^\prime$ and hadron candidates. For a track to be identified as  hadron, it was required that: its transverse
momentum was larger than 0.05~GeV/$c$, it was produced in the
current fragmentation region, as defined by the c.m. Feynman
variable $x_F\approx z-(E^h_T)^2/(zW^2)>0$, and it was
associated with a cluster in one of the hadron calorimeters  HCAL1 or HCAL2 with an energy deposit greater than 5~GeV in HCAL1 or 7~GeV in HCAL2.
 The efficiencies of the calorimetries above these energies are close to 100\%. The energy of hadrons extended up to 120 GeV in the former and up to 140 GeV in the latter. All hadrons of the SIDIS candidates were included in the analysis of asymmetries. For the final selection of the SIDIS events and hadrons, the SIDIS candidates have to pass stability tests, as described in Section 3.3 and in Section 3.4.
The total number of hadrons in 2006 after afore-mentioned selections is $15.6\times 10^6$ including $8.6\times10^6$ $h^+$ and $7.0\times10^6$ $h^-$.

To summarise, the SIDIS events and hadrons have been selected from preselected candidates requiring: 140 GeV/c $< p_\mu <$ 180 GeV/c, $Q^2 >$ 1 (GeV/c)$^2$,
0.1 $<y<$ 0.9, 0.01 $<z<$ 1, $p^h_T >$ 0.05, $x_F >$ 0, $E_{HCAL1} >$ 5 GeV,
$E_{HCAL2} >$ 7 GeV. In order to test the influence of the stronger $p^h_T$ and $z$ kinematic cuts on
the ``integrated" azimuthal asymmetries, they were calculated summing up all selected hadrons (see Section 3.6) at variance with Ref. \cite{Alekseev:2010dm}. Azimuthal asymmetries as functions of the kinematic variables $x$, $z$ or $p^h_T$ were calculated in a restricted region following Ref. \cite{Alekseev:2010dm}, i.e. summing up hadrons within the intervals given in Table 1 below. The number of hadrons within these intervals is reduced by a factor of about two compared to the total number.
\begin{table}[htb]
\centering 
\caption{\footnotesize Intervals of $x$, $z$, $p^h_T$ and their weighted mean values for which
asymmetries as functions of kinematic variables were calculated. The $Q^2$-intervals corresponding to the $x$-intervals are shown for reference.}
\begin{tabular}{|p{95pt}|p{85pt}||p{95pt}||p{95pt}|}
\hline \hfil$x$ \par intervals ~\hfill mean&\hfil $Q^2$
(GeV/$c$)$^{2}\vphantom{^{2\strut}}$\par intervals \hfill mean&\hfil $z$\par intervals
\hfill mean&
\hfil$p^{h}_{T}$ (GeV/$c$) \par intervals \hfill mean \\
\hline 0.004 -- 0.012,\hfill 0.010 \par 0.012 -- 0.022,\hfill
0.020 \par 0.022 -- 0.035,\hfill 0.031 \par 0.035 -- 0.076,\hfill
0.053 \par 0.076 -- 0.132,\hfill 0.098 \par 0.132 -- 0.700,\hfill
0.190& 1.0 -- 3.0,\hfill 1.45 \par 1.0 -- 6.0,\hfill 2.07 \par
1.0 -- 9.5,\hfill 2.89 \par 1.0 -- 20.0,\hfill 4.82 \par 2.0 --
35.0,\hfill 9.20 \par 3.0 -- 100.0,\hfill 21.26& 0.200 --
0.234,\hfill 0.216 \par 0.234 -- 0.275,\hfill 0.253 \par 0.275 --
0.327,\hfill 0.299 \par 0.327 -- 0.400,\hfill 0.361 \par 0.400 --
0.523,\hfill 0.455 \par 0.523 -- 0.900,\hfill 0.661& 0.100 --
0.239,\hfill 0.177 \par 0.239 -- 0.337,\hfill 0.289 \par 0.337 --
0.433,\hfill 0.385 \par 0.433 -- 0.542,\hfill 0.485 \par 0.542 --
0.689,\hfill 0.610 \par 0.689 -- 1.000,\hfill 0.814  \\ \hline
\end{tabular}
\label{tab1}
\end{table}

\begin{figure}[h!]
\centering
\includegraphics[width=1\textwidth]{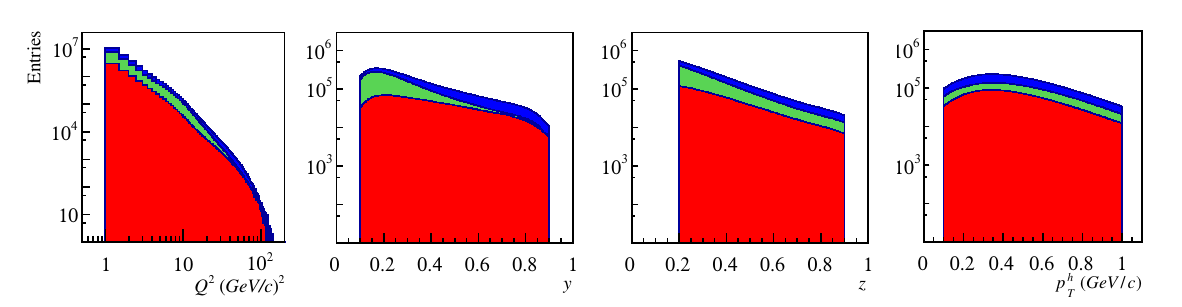}
\caption{\label{fig2}
Kinematic distributions of selected SIDIS events vs. $Q^2$ and $y$ and of charged hadrons vs. $z$ and $p^h_T$ within the region shown in Table \ref{tab1}: 2006 (lower, red), 2002 -- 2004 (middle, green) and 2002 -- 2006 (upper, blue).}
\end{figure}

The distributions of selected SIDIS events as a function of $Q^2$ and $y$ and of charged hadrons as a function of  $z$ and $p^h_T$ for different data samples are presented in Fig.~\ref{fig2}.
\subsection{Tests of data stability}
Taking advantage of the three-cell polarised target, stability
tests for the 2006 data were performed by investigating
variations of events from run to run for certain observables via ratios
$R_i$, where $i$ is the run number, using the combined
information from cells \textit{U} and \textit{D} denoted by
$(U+D)$, and that of cell \textit{M}. One expects the ratio $R_i
=(U_i+D_i)/M_i$ to be independent of luminosity, close to unity
and stable from run to run. In order to confirm this expectation,
the ratios $R_{i}$ per run were obtained for the following
observables that are relevant to the selection of SIDIS events
and hadrons:  number of SIDIS events, number of tracks per SIDIS
event, number of clusters in HCAL1 (HCAL2) with $E > 5\ (7)$ GeV,
average energy of clusters in HCAL1 (HCAL2), average energy of the
associated clusters per event in HCAL1 (HCAL2) and
average angle {$\langle\phi\rangle$}. The $R_{i}$ values as a function of the
run number were fitted by  constants $\overline{R}$ for all runs.
It was found that most of these $R_{i}$ were stable within the
$\pm 3\sigma$ limits around the average values
$\overline{R}\approx 1.05$, except for some runs and one of
the periods.

The stability of the measurement of the hadron azimuthal angle
$\phi$ in the range $\pm 180^\circ$ is essential for determination of asymmetries.
Distributions of $\phi$-values in this range were obtained for each run of data taking and average values $\langle\phi\rangle_i$ per run determined. The distribution of $\langle\phi\rangle_i$ had a Gaussian shape around the mean value equal to zero for almost all runs.
\vskip-3mm

\subsection{Acceptance-cancelling method for calculations of cross section ratios}
The modified "acceptance-cancelling" double ratio
method was used to calculate the ratios of the SIDIS cross sections for positive and negative target polarisations denoted as $\sigma_+/\sigma_-$ and the 2006 asymmetries. In this Paper, the modified double ratio method is applied in three forms: first, in a form of "acceptance-cancelling", second,  in a form of "cross section-cancelling" and, third (see Section 5), again in the "acceptance-cancelling" form to test the hadron yield stability.

In order to cancel acceptances, the method utilises double ratios,
i.e. the product of two ratios of events.
For the three-cell target the method was modified as follows. The
target cell $M$ was artificially divided in two sub-cells $M1$
and $M2$, each 28~cm long, and two pairs of cells ($U$ and $M1$)
and ($M2$ and $D$) are considered below. The cells in each pair
have equal lengths, i.e. equal densities of deuterons,
but opposite polarisations $p$ ($+$ or $-$) at a given solenoid
field direction $f$ ($+$ or $-$). For each pair of cells at a given $f$,
one can construct the double ratio using the number of selected SIDIS events or hadrons.
These numbers obtained from the cell $i$
and denoted as $N^i_{pf}$ are usually expressed via a product of a cell luminosity ($L^i_f$) given by the beam intensity times the target cell material density a target cell acceptance ($A^i_f$) and the corresponding cross section ($\sigma_p$): $N^i_{pf} = L^i_f\times A^i_f\times \sigma_p$,
i.e. luminosity, acceptance and cross section are folded in the number of events.
Taking this relation into account  as well as the COMPASS procedure
of measurements divided in two groups of runs G1 and G2, one can construct for each pair of the target cells the "acceptance-cancelling" double ratio of events (hadrons) that provides a way to unfold the $(\sigma_+/\sigma_-)^2$. Particularly, for the polarisation settings at $f = +$, the two double ratios
of event numbers constructed for the ($U,M1$) and for the ($M2,D$) pairs have the forms given by Eqs. (\ref{eq5}), where events for the first (second) ratio of each pair are taken from the G1 (G2) runs.
\begin{equation}
\left[\frac{N^U_{{+}+}}{N^{M1}_{{-}+}}\right]_{\rm G1}
\times~\left[\frac{N^{M1}_{{+}+}}{N^U_{{-}+}}\right]_{\rm G2}=
\left(\frac{\sigma_+}{\sigma_-}\right)^2_1, \qquad
\left[\frac{N^D_{{+}+}}{N^{M2}_{{-}+}}\right]_{\rm G1}
\times~\left[\frac{N^{M2}_{{+}+}} {N^D_{{-}+}}\right]_{\rm G2}=
\left(\frac{\sigma_+}{\sigma_-}\right)^2_2.
\label{eq5}
\end{equation}
Substituting in the left parts of Eqs. (\ref{eq5}) the above expressions for $N^i_{pf}$ one can see that, after  "cancellations" of $L^i_f$ and $A^i_f$, the double ratios of events are directly related to the cross section ratios squared. Because the luminosities of cells are equal, they contribute equally to the numerators and denominators and their cancellations are expected in each of the Eqs. (\ref{eq5}) ratios.
If the acceptances $A^i_f$ of the cells are similar at the same $f$ in the G1 and G2 groups of runs
(it is subject for tests below), they are also folded equally in the
corresponding number of events and ``cancel" in the
double ratios, i.e. it is not necessary to calculate them (see Section 5).
In the numerator of each ratio, the number of
events (hadrons) are taken from the runs with the positive
target cell polarisation, while in the denominator they are
taken from the runs with negative target polarisation. Thus under above conditions each
ratio of events (hadrons) in the left parts the Eqs.~(\ref{eq5}) is equal to the ratio $\sigma_+/\sigma_-$, which is known to be close to unity (it is subject for test below).
Hence, each double ratio in Eqs.~(\ref{eq5}), which
is equal to $(\sigma_+/\sigma_-)^2$, also have to be equal within statistical uncertainties and is expected to be close to unity.
The stability of acceptances during the G1 and G2 runs have been checked using the cross sections canceling double ratios of events (hadrons) in the forms similar to ones given in
Eqs. (11), (12) of Ref. \cite{Alekseev:2010dm} which are related to the ratios of acceptances.

Similarly, at $f=-$ the two double ratios constructed for the same target
pairs are: 
\begin{equation}
\left[\frac{N^U_{{+}-}}{N^{M1}_{{-}-}}\right]_{\rm G1}
\times~\left[\frac{N^{M1}_{{+}-}}{N^U_{{-}-}}\right]_{\rm G2}=
\left(\frac{\sigma_+}{\sigma_-}\right)^2_3,\qquad 
\left[\frac{N^D_{{+}-}}{N^{M2}_{{-}-}}\right]_{\rm G1}
\times~\left[\frac{N^{M2}_{{+}-}} {N^D_{{-}-}}\right]_{\rm G2}=
\left(\frac{\sigma_+}{\sigma_-}\right)^2_4.
\label{eq6}
\end{equation}
Because no requirements except time stabilities are imposed on the data, the Eqs. (\ref{eq5}, \ref{eq6}) are valid and can be used for calculations of either the cross section ratios in the restricted kinematic regions or in the whole available region (see Section 3.6).

Thus each of four double ratios of the events (hadrons) in Eqs.~(\ref{eq5}, \ref{eq6})
calculated with SIDIS events are related to the squared ratio of the SIDIS cross sections for
positive and negative target polarisations determined with a part
of the data. When statistically averaged, they can be used to calculate asymmetries with the whole data provided that (i) acceptances in the G1 and G2 periods are indeed stable and
equal, (ii) the values of the double ratios calculated for SIDIS events and for hadrons with polarisation settings at $f=+$ and
$f=-$ are stable and equal within statistical uncertainties.
These requirements were checked with SIDIS event candidates and hadrons and final selections of them were determined.

Altogether, the stability tests have shown that
(i) acceptances are stable and equal during the G1 and G2 groups of runs, (ii) the double ratios in
Eqs. (\ref{eq5}, \ref{eq6}) calculated with SIDIS events or hadrons are stable over the data
taking periods and contained inside the $\pm
3\sigma$ corridors around the average values which are close to unity.
In order to be accepted for analysis the average value per a given run of the acceptances, the angles $\langle\phi\rangle_i$ and the double ratio values
defined by Eqs.~(\ref{eq5}, \ref{eq6}) have to be within $\pm
3\sigma$ limits of the corresponding mean value for all runs.
Otherwise the run was rejected. The rejected runs contained about
10\% of 2006 hadrons.


\subsection{Extraction of azimuthal asymmetries in hadron production}
For the extraction of the azimuthal asymmetries
$a_{h^\pm}(\phi)$ off the cross section ratios, the distributions of the charged hadrons $h^+$
and $h^-$ were separately analysed as a function of the azimuthal
angle $\phi$ in the region from $-180^\circ$ to $+180^\circ$ divided into
10 $\phi$-bins. For both $h^+$ and $h^-$, the double ratios of the hadrons defined by Eqs.~(\ref{eq5}, \ref{eq6}),
$(\sigma_+/\sigma_-)^2_k(\phi)$,~ $k=1,...4$, were calculated and
combined as follows:
\vskip-3mm%
\begin{equation}
\label{eq11}
\left(\frac{\sigma_+}{\sigma_-}\right)_{h^\pm}^2\!\!\!\!\!\!(\phi)\!\!
=\!\!\!
\left[\left(\frac{\sigma_+}{\sigma_-}\right)^2_1\!\!\!\!\!\!\oplus
\!\! \left(\frac{\sigma_+}{\sigma_-}\right)^2_2\!\!\!\!\!\!\oplus
\!\! \left(\frac{\sigma_+}{\sigma_-}\right)^2_3\!\!\!\!\!\!\oplus
\!\! \left(\frac{\sigma_+}{\sigma_-}\right)^2_4\right]_{h^\pm}
\!\!\!\!\!\!\!\!\!\!(\phi) \!\!\cong\!\! 1+a_{h^\pm}(\phi)\!
\sum_{k}\!\!\left[\sum_{i,f,p\in k}{\cal
P}^i_{pf}(x)\right]_{\!\!\!h^\pm}\!\!\!\!\!\!\!\!\cdot W_k~,
\end{equation}
where the symbol $\oplus$ means statistically weighted averaging.
As it was shown in Ref.~\cite{Alekseev:2010dm}, in first
approximation, the squared ratios of cross sections $(\sigma_+/\sigma_-)^2_{h^\pm}(\phi)$
are related to the asymmetries $a_{h^\pm}(\phi)$ multiplied by polarisation
terms. For each hadron charge, the polarisation term is given by
the sum of the ${\cal P}^i_{pf}(x)$  values, each of them being the
product of target cell polarisations $|P^i_{pf}|$ and dilution
factor f$^{\,i}(x)$, as defined in Refs. \cite{Alekseev:2010dm,Ageev:2005gh},
where $i$, $p$ and $f$ are those used to calculate
the ratio $(\sigma_+/\sigma_-)^2_k(\phi)$, i.e.\ four polarisation
values at each $k$. The weight $W_k$ is equal to the ratio of the
number of hadrons, $N_k$, to the total number of hadrons, $N_{\rm
tot}$. Therefore, the $a_{h^\pm}(\phi)$,
referred to as single-hadron asymmetries, are:
\vskip-5mm
\begin{equation}
\label{eq12} a_{h^\pm}(\phi)\cong
{\large
\frac{(\frac{\sigma_+}{\sigma_-})_{h^\pm}^2(\phi)-1}{\sum\limits_{k}
[\!\!\!\sum\limits_{~i,f,p\in k}\!\!\!\!{\cal
P}^i_{pf}(x)
]_{h^\pm}\!\cdot W_k}}.
\end{equation}
\vskip-5mm
\subsection{The 2006 asymmetries}
Following Section 3.2, the asymmetries $a_{h^\pm}(\phi)$ were calculated
in this Paper (i) as the ``integrated"
asymmetries using the total number of $h^+$ or $h^-$, and (ii)
as the asymmetries vs. one of kinematic variables
$x$, $p^h_T$ or $z$ disregarding the others and using the numbers of
$h^+$ or $h^-$ within the intervals defined in Table 1.
In each case, the asymmetries were fitted by the function from Eq. (4)
using the standard least-square-method and extracting all asymmetry
modulation amplitudes simultaneously.

\paragraph{Integrated asymmetries.}
These asymmetries for the 2006 data as a function of the azimuthal angle $\phi$ are shown in
Fig.  \ref{fig1a} together with results of the fits
given in Table 2.
\begin{figure}[h!]
\label{fig1a}
\centering\vskip-2mm
\includegraphics[width=0.35\textwidth]{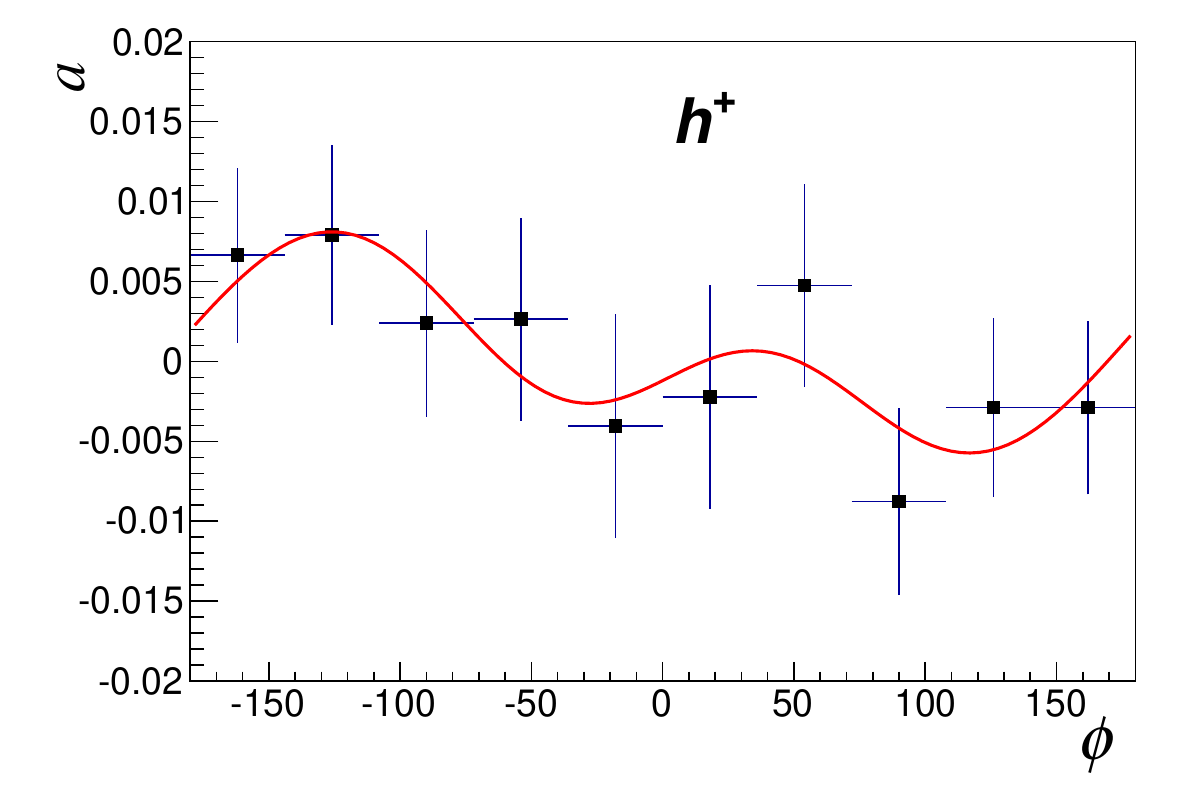}
\includegraphics[width=0.35\textwidth]{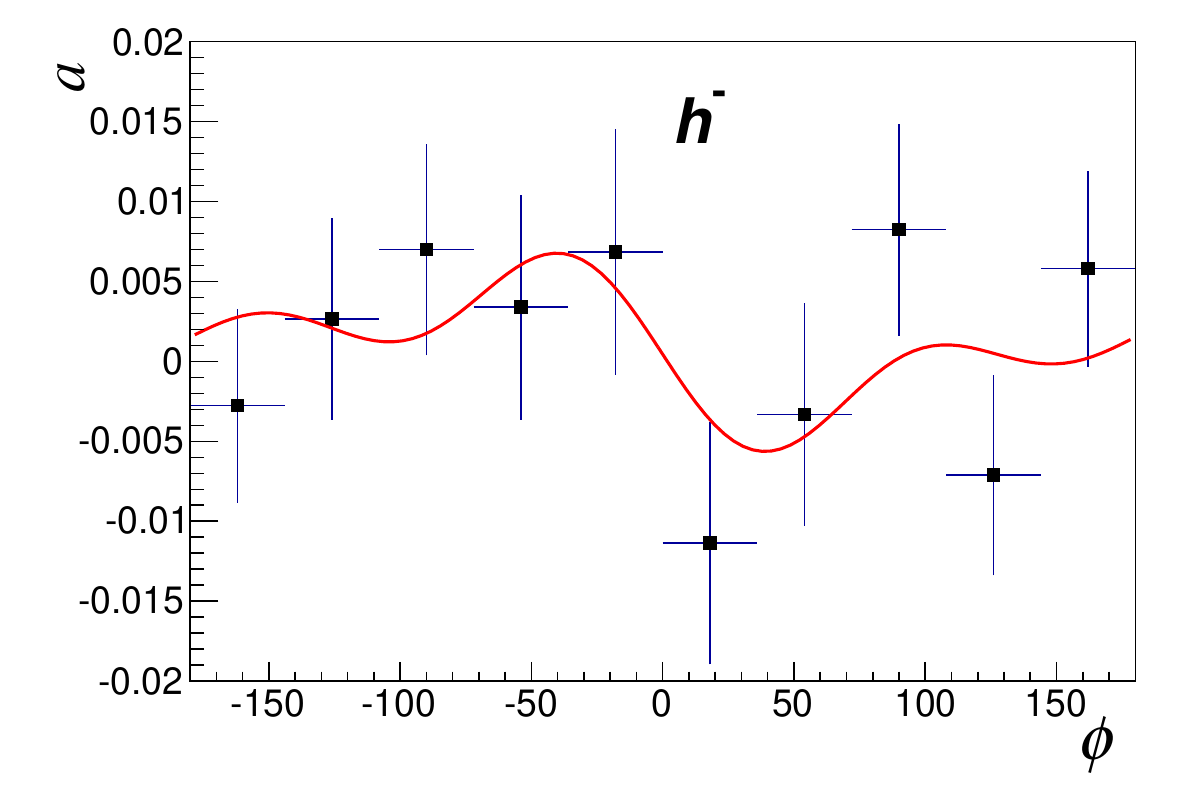}
\vskip-3mm
\caption{\label{fig1a}\footnotesize%
The 2006 ``integrated" asymmetries $a$ as functions of the azimuthal angle $\phi$.
Curves show the corresponding fits.}
\vskip-4mm
\end{figure}
\begin{table}[h!]
\begin{center}
\vskip-3mm
\caption{{\footnotesize
The $h^+$ and $h^-$ modulation amplitudes of the integrated azimuthal asymmetries obtained from the statistically combined amplitudes of the 2002--2006 (left), those of 2002--2004 (middle) and those of  2006 (right).}
}
\label{tab3}
\footnotesize
{\begin{tabular}{|c|c|c|c|c|c|c|}\hline
Modulation &\multicolumn{2}{|c|}{2002--2006 amplitudes
in $10^{-3}$}
& \multicolumn{2}{|c|}{2002--2004 amplitudes
in $10^{-3}$}&
\multicolumn{2}{|c|}{2006 amplitudes
in $10^{-3}$} \\
amplitudes
&\multicolumn{2}{|c|}{~~~~$h^+$~~~~~~~~~~~~~~~~~~~~~~$h^-$}
&\multicolumn{2}{|c|}{~~~~$h^+$~~~~~~~~~~~~~~~~~~~~~~$h^-$}
&\multicolumn{2}{|c|}{$h^+$~~~~~~~~~~~~~~~~~~~~~~$h^-$}\\
\hline
$a^{0}\vphantom{a^{0\strut}}$&$\phantom{-}2.81\pm0.96$&$\phantom{-}2.01\pm0.98$
&$\phantom{-}3.50\pm1.10$&$\phantom{-}2.30\pm1.10$
&$\phantom{-}0.35\pm1.92$&$\phantom{-}0.98\pm2.13$\\
$a^{\sin\phi} $&$-1.93\pm1.31$&$-0.74\pm1.41$
&$-1.30\pm1.50$&$-0.10\pm1.60$
&$-4.13\pm2.66$&$-2.97\pm2.98$\\
$a^{\sin2\phi}$&$-0.29\pm1.33$&~\phantom{-}$1.00\pm1.43$
&$-1.50\pm1.50$&~\phantom{-}$2.00\pm1.60$
&$\phantom{-}3.78\pm2.71$&$-2.42\pm3.00$\\
$a^{\sin3\phi}$&\phantom{-}$0.34\pm1.36$&$-0.10\pm1.42$
&~~\phantom{-}$0.30\pm1.50$&~~\phantom{-}$0.60\pm1.60$
&$\phantom{-}0.41\pm2.69$&$-2.42\pm3.01$\\
$a^{\cos\phi} $&\phantom{-}$1.52\pm1.32$&$\phantom{-}0.66\pm1.42$
&~~\phantom{-}$2.40\pm1.50$&~\phantom{-}$1.00\pm1.60$
&$-1.58\pm2.74$&$-0.53\pm3.03$\\
\hline
\end{tabular}}
\normalsize
\end{center}
\vskip-3mm
\end{table}

\begin{figure}[h!]
\includegraphics[width=1\textwidth]{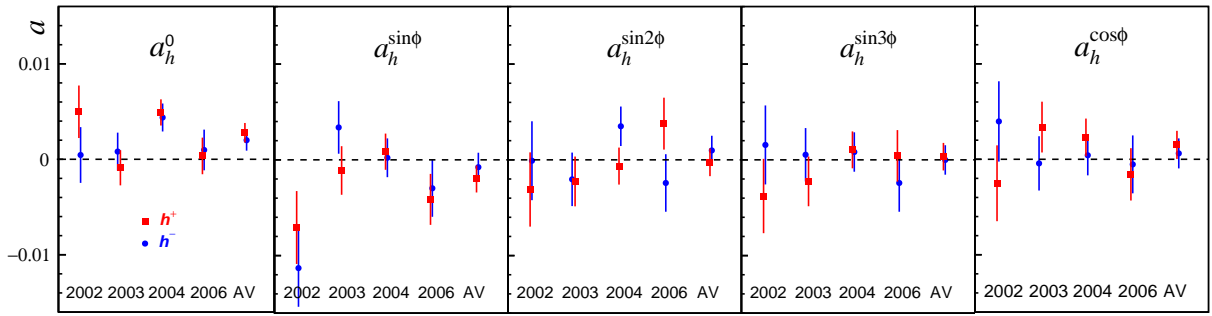}
\vskip-3mm
\caption{\footnotesize
The values of modulation amplitudes $a$ together with their uncertainties obtained from the fits of the integrated asymmetries $a_{h^\pm }(\phi)$ by the function
from Eq.~(\ref{eq3}) separately for the data of 2002, 2003, 2004
and 2006 as well as statistically combined modulation amplitudes for four years denoted by AV (see Section 4.1).}
\label{fig3}
\vskip-3mm
\end{figure}

In order to compare the 2006 integrated asymmetries
to those of the 2002, 2003 and 2004, the latter ones were recalculated using the
total number of hadrons. The results of the fit
for the 2006 integrated asymmetries together with those of the
2002, 2003 and 2004 calculated similarly are shown in Fig. \ref{fig3}. The modulation
amplitudes obtained for each year are in agreement with one another, as confirmed by the compatibility tests (see Section 5).

In order to check the impact of the disregarded modulations,
which could have appeared from SIDIS partial cross sections
d$\sigma_{00}$ and d$\sigma_{L0}$, on the modulation amplitudes
in Eq.~(\ref{eq3}), we have performed fits of the 2006 integrated asymmetries by a new fitting function which contains a numerator and denominator. In the numerator we have used the same modulations
as in Eq.~(\ref{eq3}), but in the denominator we included the disregarded
modulation with average amplitudes determined in
Ref.~\cite{Adolph:2014pwc}. For the asymmetry $a_{h^-}(\phi)$, the fitting function is as follows:
\vskip-6mm%
\begin{equation}
\label{eq15} a_{h^-}(\phi)=\frac{a_{h^-}^{0} + a_{h^-
}^{\sin\phi}\sin\phi+ a_{h^-}^{\sin2\phi}\sin2\phi +
a_{h^-}^{\sin3\phi}\sin3\phi +
a_{h^-}^{\cos\phi}\cos\phi}{1+0.0412\cos2\phi+
0.0552\cos\phi+0.0008\sin\phi}~.
\end{equation}
\vskip-3mm%
By comparing results of this fit with results of the standard fit of
the 2006 data shown in Fig.~\ref{fig3}, it was found that the
differences between values of modulation amplitudes in the
numerator are smaller than 1\% of the fit uncertainties.
Similar results are obtained for $a_{h^+}(\phi)$ replacing
amplitudes in the denominator of Eq.~(\ref{eq15}) by
corresponding values from Ref.~\cite{Adolph:2014pwc}. Thus the
contributions of the disregarded modulations to the integrated asymmetries in Eq.~(\ref{eq3}) are indeed negligible.

\paragraph{Asymmetries as functions of kinematic variables.}
The 2006 modulation amplitudes as  functions of kinematic
variables were compared to those from the combined 2002 -- 2004
data and found to be in agreement within uncertainties of the fits.
They are used for calculations of the combined 2002-- 2006 modulation amplitudes (see Section 4.2).

\section{Azimuthal asymmetries for the combined deuteron data}
\subsection{Integrated asymmetries}
For the integrated asymmetries, the values of the combined 2002 -- 2006 modulation amplitudes, which are shown in Fig. \ref{fig3} and denoted by AV, were obtained using a statistical combination of four corresponding amplitudes.
They are given in Table~\ref{tab3}. The combined 2002 -- 2004 modulation amplitudes, calculated as for 2006, are also shown in Table \ref{tab3} in order to allow comparison to those of Ref. \cite{Alekseev:2010dm}, which were calculated in the restricted kinematic  region. As expected for the iso-scalar deuteron target, consistent results
are obtained for the $\phi$-independent terms $a_{h^+}^{0}$ and
$a_{h^-}^{0}$. All $\phi$-modulation amplitudes are consistent
with zero within uncertainties of fits. Comparing results for the 2002 -- 2006 combined data presented in Table \ref{tab3} to the results for 2002 -- 2004 and to the results of Ref. \cite{Alekseev:2010dm}, one can see that they are in agreement between themselves within the quoted uncertainties. This indicates that the integrated asymmetries with and without kinematic cuts of Table \ref{tab1} are consistent, i.e. these cuts reduce statistics but do not change the values of the asymmetries within experimental uncertainties.
Due to increased statistics of each year, the statistical uncertainties of the combined 2002 -- 2006 amplitudes are reduced by a factor of about 1/1.6 compared to those of Ref. \cite{Alekseev:2010dm}.
\subsection{Asymmetries as functions of kinematic variables}
\begin{figure}[h!]
\centering 
\includegraphics[width = 1\textwidth]{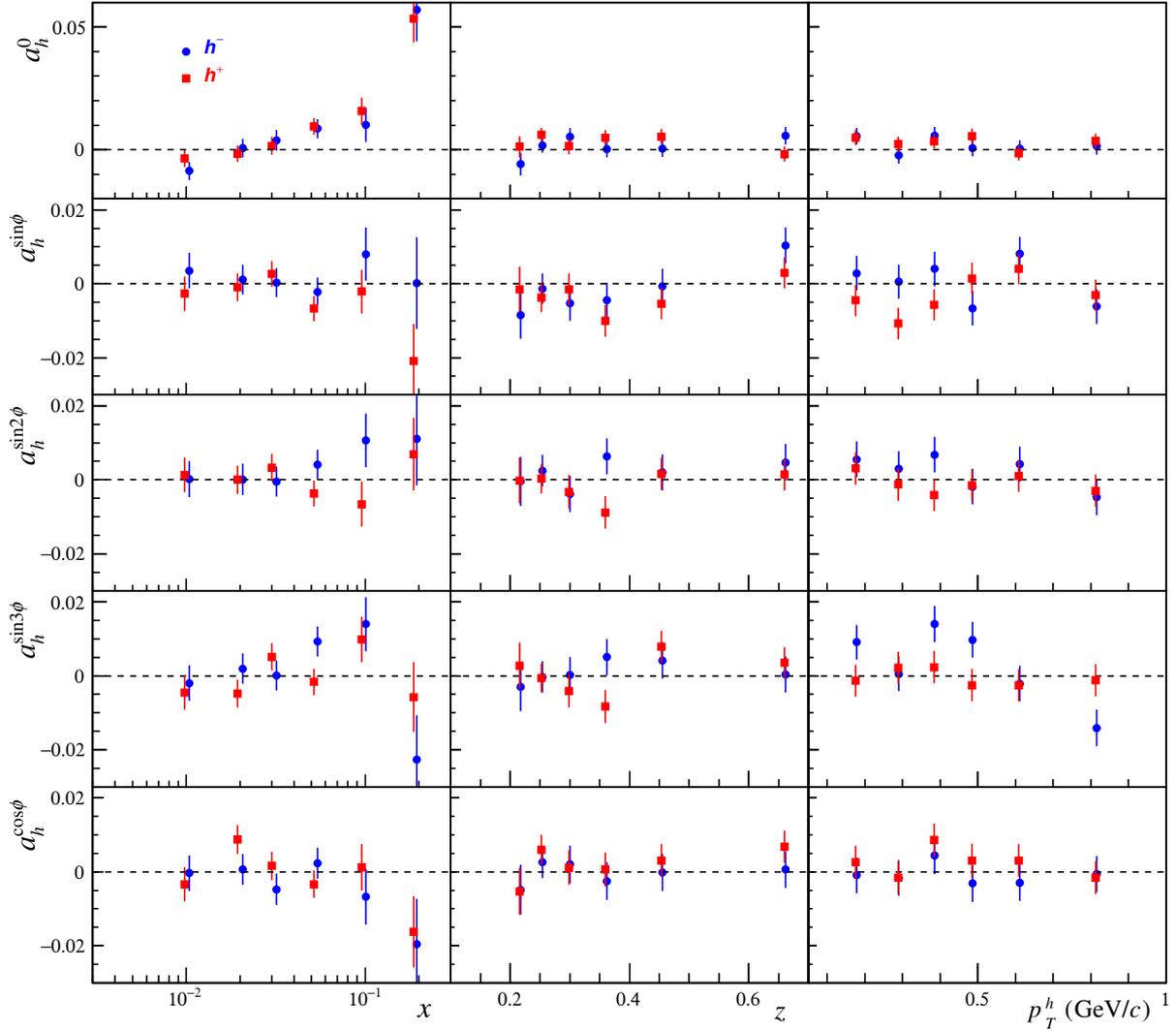}
\vskip-2mm
\caption{\label{fig4}\footnotesize
The modulation amplitudes $a$ of the $h^+$ and $h^-$ azimuthal
asymmetries as the function of $x\ ,z$ and $p^h_T$ obtained from
the combined 2002 -- 2006 data on the muon SIDIS off longitudinally
polarised deuterons. Only uncertainties of fits are shown.}
\end{figure}

The final 2002 -- 2006 results on the modulation amplitudes of
asymmetries $a_{h^\pm}(\phi)$ calculated as the function of one of
the variables $x,\ z$ and $p^{h}_{T}$ while disregarding the
others were obtained from the statistically averaged 2002, 2003,
2004 and 2006 modulation amplitudes. The results are presented in Fig.~\ref{fig4}.

\begin{figure}[h!]
\centering 
\includegraphics[width=.5\textwidth,bb=64 9 541 414]{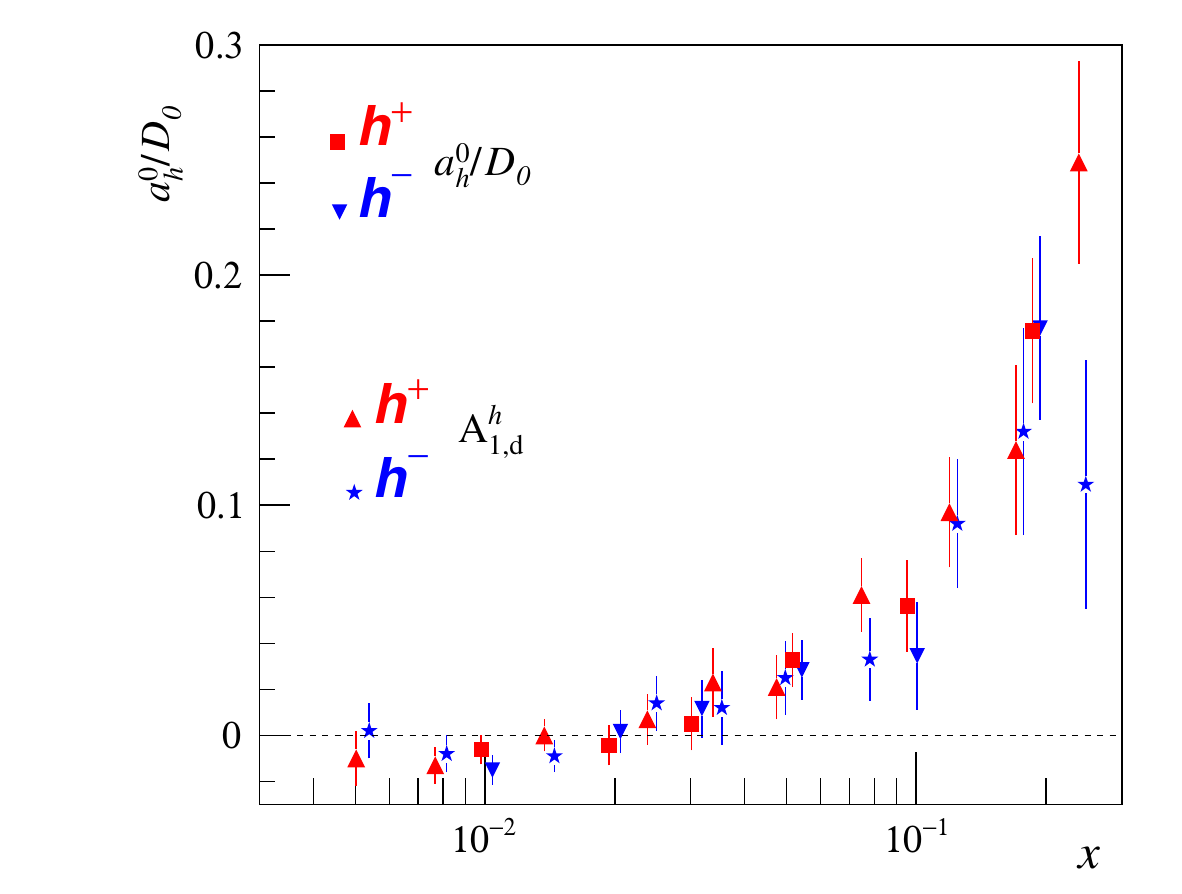}
\vskip-2mm
\caption{\label{fig5}
The $x$-dependence of the values $a_{h^\pm }^0(x)/D_{0}(x,y)$ for
2002 -- 2006 data compared to the data of Ref.~\cite{Alekseev:2007vi}
on the asymmetries $A^{h^\pm}_{1d}(x)$.} 
\end{figure}

Except for the $a^{0}_{h^\pm}(x)$, all
amplitudes when fitted by constants are found to be consistent
with zero within statistical uncertainties ($\chi^2/NDF\simeq 1$).
As expected, the  $a_{h^\pm }^0(x)$ for deuteron target have
the same $x$-de\-pendence for positive and
negative hadrons. Addi\-tionally, the $x$-de\-pen\-dence of the
$a^0_{h^\pm}(x)/D_0(x,y)$ values are presented in
Fig.~\ref{fig5}, where $D_0(x,y)$ is the virtual-pho\-ton depolarisation
factor for each $x$ interval multiplied by the average beam polarisation $|P_\mu|$, as
defined in Ref.~\cite{Alekseev:2010dm}. If the $a_{h^\pm}^0(x)$
represent the main contributions to the asymmetries of Eq.~(\ref{eq2}), the values of
$a_{h^\pm}^0(x)/D_{0}(x,y)$ by definition (see e.g.\ Ref.~\cite{Anselmino:1994gn})
should be equal to the asymmetries
$A^{h^\pm}_{1d}(x)$. Within experimental uncertainties, there is
a good agreement between our data on $a_{h^\pm }^0(x)/D_{0}(x,y)$
and the data of Ref.~\cite{Alekseev:2007vi} on
$A^{h^\pm}_{1d}(x)$, which confirms the correctness of the
results on the asymmetries calculated by the modified
acceptance-cancelling method. The values of $A^{h^\pm}_{1d}(x)$
were obtained with the 2002 -- 2004 data. A similar $x$-dependence
was also observed with 2002 -- 2006 data for the asymmetries
$A^{\pi^\pm}_{1d}(x)$ and $A^{K^+}_{1d}(x)$ obtained with the
identified hadrons (see Ref.~\cite{Alekseev:2009ac}).
\section{Systematic uncertainties}
The global compatibility test of the results on the asymmetries
$a_{h\pm}(\phi)$, that were obtained separately for 2002, 2003,
2004 and 2006 years, was performed by building the pull
distribution: ${pull}_i=(a_i-\langle
a\rangle)\times|\sigma_{a_i}^2- \sigma_{\langle
a\rangle}^2|^{-1/2}$, where $a_{i}$ is the asymmetry for a given
year, hadron charge and kinematic interval, $\langle a\rangle$ is the
corresponding weighted mean value over four years and $\sigma$
denotes the corresponding standard deviation. The pull distribution
had in total 750 entries compared to 540 for 2002 –- 2004 years.
The asymmetries reported in this Paper, in principle, could have the non-cancelled-acceptance or luminosity time-dependent effects folded in the event numbers of one of the Eqs. (\ref{eq5}, \ref{eq6})-ratio and, consequently, in one or several values of $a_i$ distorting the pull distribution. In the absence of such effects, as expected, the pull distribution follows the Gaussian distribution with the mean value consistent with zero ($-0.038\pm 0. 033$ in our case) and $\sigma$ with unity (0.978$\pm$0.024). This indicates that no significant time-dependent systematic effects are present in the data on asymmetries, i.e. systematic uncertainties in values of $a_i$  are smaller than statistical ones.

\begin{figure}[h!]
\centering
\includegraphics[width=0.325\textwidth]{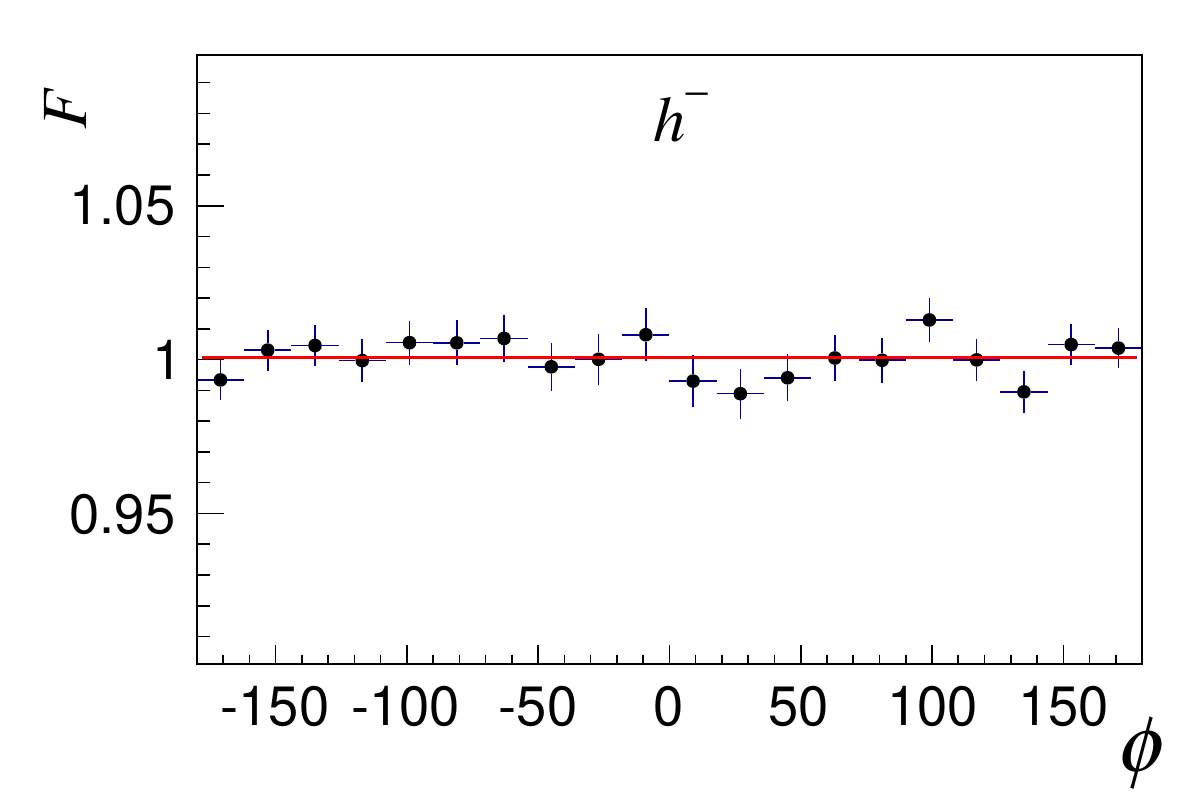}
\includegraphics[width=0.325\textwidth]{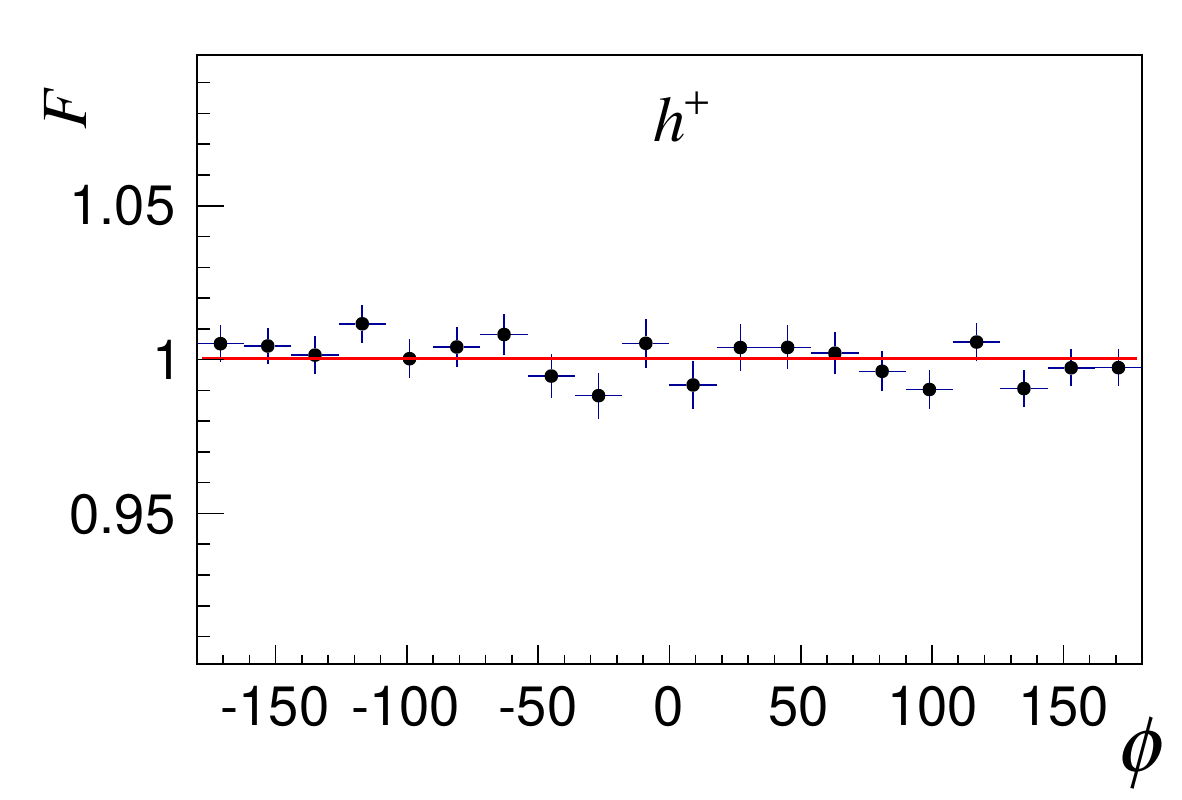}
\includegraphics[width=0.325\textwidth]{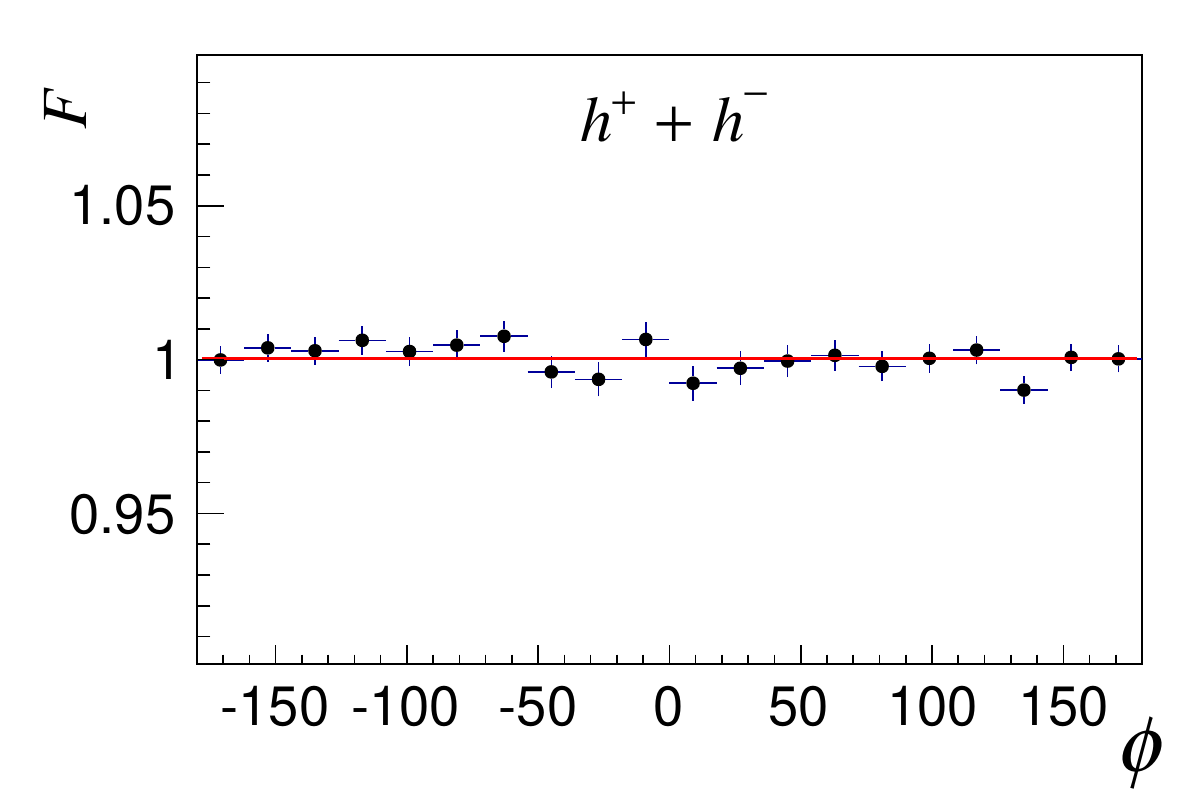}
\caption{\footnotesize%
\label{fig7}The $\phi$-dependence of the weighted sums of double
ratios $F(\phi)$ for 2006 data: $h^-$, $h^+$ and $h^++ h^-$.  The solid (red) lines represent the results of fits by  constants.}
\end{figure}

Quantitative measures of possible systematic effects have been obtained by estimating additive and multiplicative uncertainties.
Main contributions to possible additive systematic  uncertainties could come from the instabilities of the hadron yields.
The $\phi$-stability of hadron yields in the 2006 data was checked following the procedure described in Ref.~\cite{Alekseev:2010dm}. For this purpose, the double ratios
of hadron numbers as a function of the azimuthal angle $\phi$ for
different polarisation settings at the field $f$ during the G1 and
G2 runs were calculated as follows:
\begin{equation} \label{eq13}
f=+:~~ F_+(\phi)=\frac{N_{{+}+}^U +
N_{{+}+}^{D}}{N_{{-}+}^{M}}\cdot
\frac{N_{{+}+}^M}{N_{{-}+}^U+N_{{-}+}^D}~, \quad
f=-:~~ F_-(\phi)=\frac{N_{{+}-}^U + N_{{+}-}^{D}}{N_{{-}-}^{M}}
\cdot \frac{N_{{+}-}^M}{N_{{-}-}^U+N_{{-}-}^D}~.
\end{equation}
Here,~ $N_{{p}f}^{i}$ is the number of hadrons per $\phi$-bin from target cell $i$ with polarisation $p$ and field $f$, as explained in Section 3.4. The ratios given by Eqs. (\ref{eq13}) are modifications of ratios used in
Ref.~\cite{Alekseev:2010dm} for the case of two target cells.
In the above double ratios, we expect the acceptance and luminosity cancellations and, as a result, the $\phi$-stability of the hadron
yields. If unstable, they could indicate possible
systematics in the acceptance as well.  The fits
by constants (see Fig. \ref{fig7}) of the weighted sums $F(\phi)=F_+(\phi)\oplus
F_-(\phi)$ in the $\phi$-region from $-180^\circ$ to $+180^\circ$
for $h^+$, $h^-$ and  $h^+ +h^-$ of the 2006 data gave
results consistent with unity within statistical uncertainties of
the order of 0.001. This means that no $\phi$-instabilities and acceptance-changing (not cancelled) effects have been observed, i.e. there are no large additive systematic uncertainties in the 2006 data. The value $\Delta a_{h^\pm}(\phi)=\pm
0.001$  was chosen as a quantitative measure of possible additive systematic uncertainties in the 2006
asymmetry measurements. It is equal to $\pm \sigma$ of the
$h^{+}+h^{-}$ data stability test for $F(\phi)$. The same value was obtained for the 2002 -- 2004 data \cite{Alekseev:2010dm} and hence adopted also for the combined 2002 -- 2006 deuteron data.

Possible sources of multiplicative systematic uncertainties of the asymmetry evaluation
are uncertainties in the determination of the beam and target polarisations and
estimations of the dilution factor.
The multiplicative systematic uncertainties of the
extracted asymmetries due to uncertainties in the determination
of the beam and target polarisations were estimated to be less
than 5\% each and those due to uncertainties of the dilution
factor to be less than about 2\%. When combined in quadrature,
overall possible multiplicative systematic uncertainty of less than 6\%
was obtained.

\section{Conclusions}
The searches for possible azimuthal modulations in the
single-hadron azimuthal asymmetries
$a_{h^\pm}(\phi)$, as a manifestation of  TMD PDFs describing the
nucleons in the longitudinally polarised deuteron, have been
performed using all COMPASS deuteron data and the
acceptance-cancelling method of analysis. For each hadron charge,
beside the $\phi$-independent term, four possible modulations
predicted by theory ($\sin\phi,\ \sin2\phi,\ \sin3\phi$ and
$\cos\phi$) and their dependence on kinematical variables are
considered. The asymmetries have been calculated both for all selected hadrons ("integrated" asymmetries) and for hadrons  as functions of kinematic variables within the restricted region.

For the "integrated" asymmetries, it was found that results in the restricted
range of kinematic variables are consistent with those of the wider range. In other words, the restricting of the kinematic region reduces the statistics but does not change the values of the asymmetries beyond the sensitivity of this experiment. The same result was obtained in Ref. \cite{Alekseev:2010dm} for the asymmetries as a function of $z$.

The $\phi$-independent terms $a_{h^\pm}^0(x)$ of the asymmetries
$a_{h^\pm}(\phi)$, which are expected to originate mostly from
the known helicity PDFs $g_{1L}(x)\equiv g_1(x)$, are connected to the virtual
photon asymmetry $A^{h^\pm}_{1d}(x)=a_{h^\pm}^{0}(x)/D_0(x,y)$.
There is good agreement between the COMPASS data on
$a^{0}_{h^\pm}(x)/D_0(x,y)$ and $A^{h^\pm}_{1d}(x)$ from Refs.
\cite{Alekseev:2009ac,Alekseev:2007vi} which confirms this
expectation.

No statistically significant dependences of
$\phi$-modulation amplitudes 
were observed as functions of $x$, $z$ or $p^h_T$ when
fitted by constants. Still, there
are some hints (statistically not confirmed) for a possible $x$-dependence of the $\sin2\phi$, $\sin3\phi$ and $\cos\phi$ modulation amplitudes. The $\sin2\phi$ amplitude for $h^{-}$
is mostly positive and rises with increasing $x$, while for
$h^{+}$ it is mostly negative and decreases with $x$. This
behaviour agrees with that discussed in Refs.
\cite{Avakian:2016rst,Airapetian:2002mf,Avakian:2007mv}, if one takes into account
different definitions of asymmetries
by the HERMES and COMPASS Collaborations. The increase with $x$ of the modulus of the $\cos\phi$ amplitudes, related to the Cahn-effect  \cite{Cahn:1978se} and predicted in Ref.~\cite{Anselmino:2006yc},
was already visible from the 2002 -- 2004 data \cite{Alekseev:2010dm}
and persists for the combined 2002 -- 2006
data. Hints for a possible $x$-dependence of $\sin3\phi$ modulation amplitudes are discussed in Ref. \cite{Avakian:2016rst}. Quantitative estimates of a possible contribution of the
$\cos\phi$ modulation to the deuteron asymmetries, related to
TMD PDFs $g^\perp_L$ and $e_L$, have been obtained in
Ref.~\cite{Mao:2016hdi}. They are in agreement with our data.

Altogether, one can conclude that contributions of TMD PDFs convoluted with
FFs to the azimuthal asymmetries in the cross sections of hadron
production in muon SIDIS  off longitudinally polarised deuterons
are small. This is either  due to possible cancellations of the
contributions to the asymmetries by the deuteron's up and  down
quarks, or/and due to the smallness of the transverse component
of the target polarisation and of the suppression factor that
behaves as $M/Q$. Some of these conclusions can be checked
studying these asymmetries in muon SIDIS off longitudinally
polarised protons.\\[-7mm]

\section*{Acknowledgements}
We gratefully acknowledge the support of our funding agencies and
of the CERN management and staff and the skills and efforts of the
technicians of our collaborating institutes. Special thanks go to
V.~Anosov and V.~Pesaro for their technical support during the
installation and running of this experiment.

\end{document}